\pdfoutput=1
\documentclass[]{article}
\usepackage{mathrsfs,amsfonts,graphics,cancel,color}
\usepackage[english]{babel}

\def\bra#1{\mathinner{\langle{#1}|}}
\def\ket#1{\mathinner{|{#1}\rangle}}
\def\braket#1{\mathinner{\langle{#1}\rangle}}

{\catcode`\|=\active 
  \gdef\Braket#1{\begingroup
\mathcode`\|32768\let|\BraVert\left<{#1}\right>\endgroup}}
\def\BraVert{\egroup\,\mid\,\bgroup}

\definecolor{Blue}{rgb}{0,0,1}
\definecolor{Red}{rgb}{1,0,0}
\definecolor{Green}{rgb}{0,1,0}
\definecolor{Purp}{rgb}{.2,0,.2}
\definecolor{white}{rgb}{1,1,1}

\newcommand{\A}{\mathcal{A}}
\newcommand{\B}{\mathcal{B}}
\newcommand{\I}{\mathcal{I}}
\newcommand{\M}{\mathcal{M}}
\newcommand{\Lm}{\mathcal{L}}
\newcommand{\K}{\mathcal{K}}
\newcommand{\s}{\mathcal{S}}
\newcommand{\e}{\mathcal{E}}
\newcommand{\se}{\mathcal{SE}}
\newcommand{\rs}{\rho^{\mathcal{S}}}
\newcommand{\re}{\rho^{\mathcal{E}}}
\newcommand{\rse}{\rho^\se}
\newcommand{\tr}{\mbox{tr}}
\newcommand{\tre}{{\rm tr}_\mathcal{E}}

\newcommand{\m}{{(m)}}
\newcommand{\n}{{(n)}}
\newcommand{\mn}{{(mn)}}
\begin{document}

\title{Operational approach to open dynamics and quantifying initial correlations}

\author{Kavan Modi\\
Department of Physics, University of Oxford\\
Clarendon Laboratory, Oxford, OX1 3PU, UK\\
\\
Centre for Quantum Technologies\\
National University of Singapore, Singapore 117543\\
\\
\texttt{kavan@quantumlah.org}}

\date{\today}
\maketitle

\begin{abstract}
A central aim of physics is to describe the dynamics of physical systems. Schr\"odinger's equation does this for isolated quantum systems. Describing the time evolution of a quantum system that interacts with its environment, in its most general form, has proved to be difficult because the dynamics is dependent on the state of the environment and the correlations with it. For discrete processes, such as quantum gates or chemical reactions, quantum process tomography provides the complete description of the dynamics, provided that the initial states of the system and the environment are independent of each other. However, many physical systems are correlated with the environment at the beginning of the experiment. Here, we give a prescription of quantum process tomography that yields the complete description of the dynamics of the system even when the initial correlations are present. Surprisingly, our method also gives quantitative expressions for the initial correlation.
\end{abstract}


There is a rich history to the studies of decoherence of quantum systems due to the interactions with the surrounding degrees of freedom. When the dynamics of the system ($\s$) is Markovian it can be described by a master equation~\cite{kossakowski, lindblad, sudarshangorini}. Nowadays many researchers are interested in systems that are non-Markovian, as there is mounting evidence that some natural systems of importance may be non-Markovian~\cite{Engeletal} and such features may allow to manipulate and control quantum systems in desired ways. There is also a great deal of interest in systems that are initially correlated with their environments ($\e$) because non-Markovianity and initial system-environment ($\se$) correlations are intimately related~\cite{tannor99, geva:1380, Rodriguez08}.

Grasping the mathematical and physical aspects of non-Markovian systems, especially with initial  $\se$ correlations, has proved to be a tough road. Nevertheless, there is a great deal of progress on deciding whether a system is non-Markovian in the recent years~\cite{Wolfetal, Rodriguez08, Breueretal, Rivasetal, devi}. However, avoiding the initial $\se$ correlations is not always possible in reality~\cite{Laineetal, paris, PhysRevA.84.042113}. Working with initial correlations in practice has proved to be much trickier than in theory. This is because the presence of correlations do not allow for a clear definition of the state $\s$ independent from the state of $\e$ and vice versa. Physical systems are complicated and have many additional degrees of freedom that are not of experimental interest. Yet these extra degrees of freedom interact with the degrees of interest leading to correlations. Therefore initially uncorrelated $\se$ state is often an approximation.

In theory of open quantum systems, discrete quantum transformations are described by the dynamical map formalism~\cite{PhysRev.121.920, kraus}: $\B(\rs)=\rs_t$. The dynamical map can be thought of as coming from the contraction of $\se$ unitary dynamics. Let us write the state of $\se$ as 
\begin{equation}\label{corrmatrix}
\rho^\se = \rho^\s \otimes \rho^\e + \chi^\se,
\end{equation}
where $\chi^\se$ is the correlations matrix~\cite{Carteret08a}. The dynamical map is the mapping from the initial states of $\s$ to the final states of $\s$, resulting from unitary dynamics of the $\se$ state
\begin{eqnarray} 
\B (\rs) = \rs_t &=& \tre \left[U \rse U^\dag\right]\label{NCPmap}\\
&=& \tre \left[U \rs \otimes \re U^\dag\right]+
\tre \left[U \chi^\se U^\dag\right]\\
&=& \B^{\rm CP} (\rs) + \B^{\rm aff},\label{CPmap}
\end{eqnarray}
where $\B^{\rm CP}$ is a completely positive map and $B^{\rm aff}$ is the affine correction term due to the initial $\se$ correlations. This means that $\B$ may not a be completely positive map when $\chi^\se \ne 0$, nevertheless it fully describes the dynamics of $\s$~\cite{Shaji05a}. However, to determine such a map experimentally would require preparing different states of $\s$ while keeping the $\se$ correlations fixed. Such preparations are not operationally feasible because altering the state of $\s$ will also alter the $\se$ correlations. Therefore, a nonpositive dynamical map is not an operationally meaningful quantity.

The operational approach to quantum dynamics relies on the fact that quantum theory is a theory of preparations and measurements. The experimental method to determine a dynamical map corresponding to a quantum process is called {\it quantum process tomography} (QPT)~\cite{JModOpt.44.2455, PhysRevLett.78.390}. It is the central tool in determining a discrete quantum process; e.g. quantum gates~\cite{Nielsen:1998py, PhysRevA.64.012314, PhysRevLett.91.120402, Wein:121.13, orien:080502, NeeleyNature, chow:090502, Howard06, myrskog:013615} or chemical reactions~\cite{Pomyalov201098, Yuen-ZhouJCP, yuen-zhou}. To see the difference between QPT and dynamical maps let us review the four basic steps necessary to carried out QPT~\cite{Kuah07a, prep2}:
\begin{itemize}
\item[(i)] Input states that span the space of $\s$ are prepared. 
\item[(ii)] The input states are sent through the process. 
\item[(iii)] The corresponding output states are determined by quantum state tomography.
\item[(iv)] The knowledge of input states, the corresponding output states, and assuming linearity completely determines the process. 
\end{itemize}

\begin{figure}[t]
\resizebox{8.7 cm}{4.37 cm}{\includegraphics{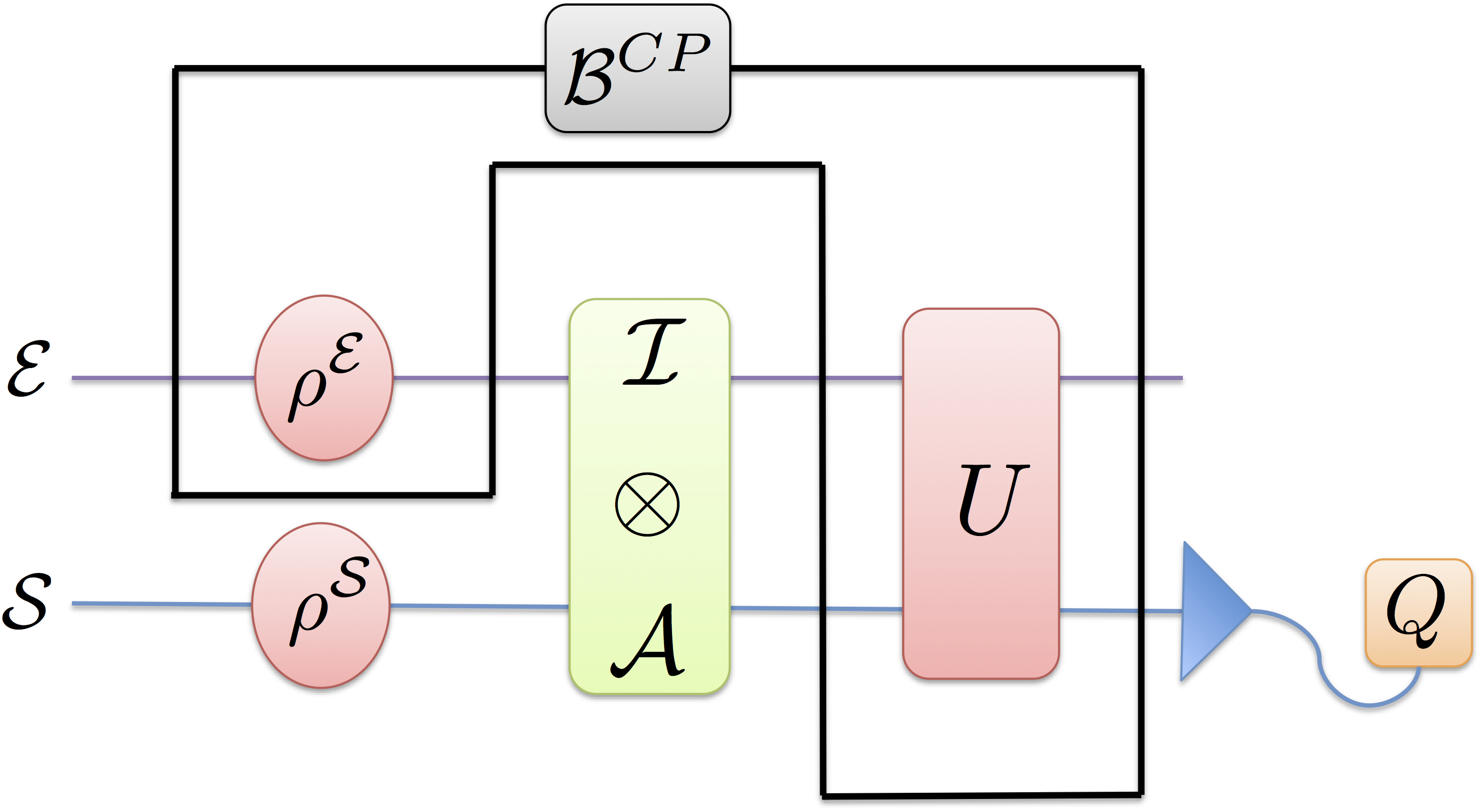}}
Figure 1: \emph{Standard quantum process tomography.} At the beginning of the experiment the system-environment state is uncorrelated. A preparation ($\A$) is made on the system and the corresponding output state $Q$ is observed. This process is described by the completely positive map of Eq.~(\ref{CPmap}), which is a function of initial state of environment and the unitary dynamics. It maps the initial states of the system to output states $Q$.
\end{figure}

Let us denote input states as $P$ and output states as $Q$. The first step of QPT is state preparation. A preparation procedure takes an unknown state of $\s$ to a known state of $\s$. Mathematically, it is described by a completely positive map acting on the system~\cite{preppap}. For instance, consider a set of preparations that project $\s$ into pure states: $[\mathcal{A}^\m \otimes \I] \left( \rse \right)= P^\m \otimes \rho^{\e|\m}$.  Since $P^\m$ is a pure state, the post-preparation $\se$ state is fully uncorrelated, where $\rho^{\e|\m}$ is the conditional state of $\e$. $\I$ is the identity operator acting on $\e$, as we assume that the preparation procedure only acts on $\s$ and not $\e$. We will discuss the implications of relaxing this assumption in Discussions. Lastly, if the preparation is not trace preserving, it should be divided $\tr \left[ \mathcal{A}^\m(\rse) \right]$ for normalisation.

The $\se$ evolution, after the preparation yields the output state:
\begin{eqnarray}
Q^\m &=& \tre \left[U [\A^\m \otimes \I](\rse) U^\dag\right] \label{PrEq}\\
 &=& \tre \left[U P^\m \otimes \rho^{\e|\m} U^\dag\right]. \label{PrepAf}
\end{eqnarray}
The key difference between the dynamical map in Eqs.~(\ref{NCPmap}) and (\ref{PrEq}) is the act of state preparation. Because dynamical maps of do accommodate state preparation, they are not operationally defined. In the presence of initial $\se$ correlations, state preparation affects the state of $\e$ in a nontrivial manner. That is, the state of $\e$ in Eq.~(\ref{PrepAf}) is conditioned by the choice of the preparations. 

In deriving the standard QPT procedure it is implicitly assumed that the initial state of $\se$ is uncorrelated~\cite{Nielsen00a}, i.e., the state of $\e$ is thought to be a constant of the problem. When that is the case, the state of $\e$ in Eq.~(\ref{PrepAf}) is not conditioned by the preparation procedure. In this case the derived map for the process is completely positive and is the same as the completely positive dynamical map in Eq.~(\ref{CPmap}). See Fig. 1 for a graphical illustration. In the presence on initial $\se$ correlation, the conditional state of $\e$ will be different for each preparation, and the assumption of linearity in step (iv) of QPT is violated, i.e. the map is a function of the preparation procedure. Such maps are nonpositive, nonlinear, or simply put nonsensical~\cite{Kuah07a, prep2}.

It then begs the question, can we determine the dynamics of a system that is initially correlated with $\e$? This is an important question for two reasons: First, there may be physical system of interest that may have initial correlations. Is it possible to study their dynamics? Second, for foundational reasons we may care to know what are the limitations in describing the dynamics of physical systems. A partial solution to these questions was given in~\cite{Kuah07a, modidis}. In this article we show that not only complete dynamics of initially correlated system can be determined, we can also determine the contribution due to initial correlations.

\section*{Results}\label{results}

\subsection*{A map on a map}

QPT is performed out by noting how input states, that span the space of $\s$, map to output states. The key insight in what follows is that it is not the input states of $\s$ that are relevant, rather it is the preparation procedures itself, i.e., the preparation map $\A$. For a $d_\s$ dimensional system there are $d_\s^2$ linearly independent states that span its space. However, there are $d_\s^4$ linearly independent operations (preparations) that span the space of preparations. If we determine the corresponding output states for a set of linearly independent preparations then by linearity we have can predict the output state for any preparation. Let us denote this map as $\M$-map.

The form of $\M$-map arises naturally when considering the whole process in physical terms: At the beginning of the experiment $\se$ is in an unknown (correlated) state, $\rse$. The system is prepared into a known input state by the preparation procedure $\A$, followed by a joint unitary dynamics. The output is given by tracing over the environmental degrees of freedom:
\begin{equation}\label{output}
Q = \tre \left[U [\A \otimes \I] \left(\rse \right) U^\dag \right].
\end{equation}
We want a map acting on the preparation map $\A$ and yielding the output state $Q$: $\M \left(\A \right) \to Q$. Then $\M$-map is everything on the right hand side of Eq.~(\ref{output}) that is not $\A$. The expression for $\M$-map in terms of matrix indices is
\begin{equation}\label{mmap}
\M_{rr'r''; ss's''} = U_{r \epsilon,r' \alpha} \rse_{r''\alpha, s''\beta} U^*_{s \epsilon,s' \beta}.
\end{equation}
Above a sum over repeated indices is implied. $\M$-map is a `super super-operator' that acts on the super operator $\A$. $\M$-map is a $d^3_\s \times d^3_\s$ tensor, which is contracted with a preparation $\A$, a $d^2_\s \times d^2_\s$ tensor, yielding the output state $Q$, a $d_\s \times d_\s$ matrix. In term of matrix indices, the action is as follows:
\begin{equation}\label{mmapaction}
Q_{rs}=\M_{rr'r''; ss's''}(\A_{r'r'';s's''}).
\end{equation}
Again, a sum over repeated indices is implied. In Methods, a full derivation for $\M$-map in the last equation is given. See Fig. 2 for a graphical illustration of $\M$-map.

\begin{figure}[t]
\resizebox{8.7 cm}{4.27 cm}{\includegraphics{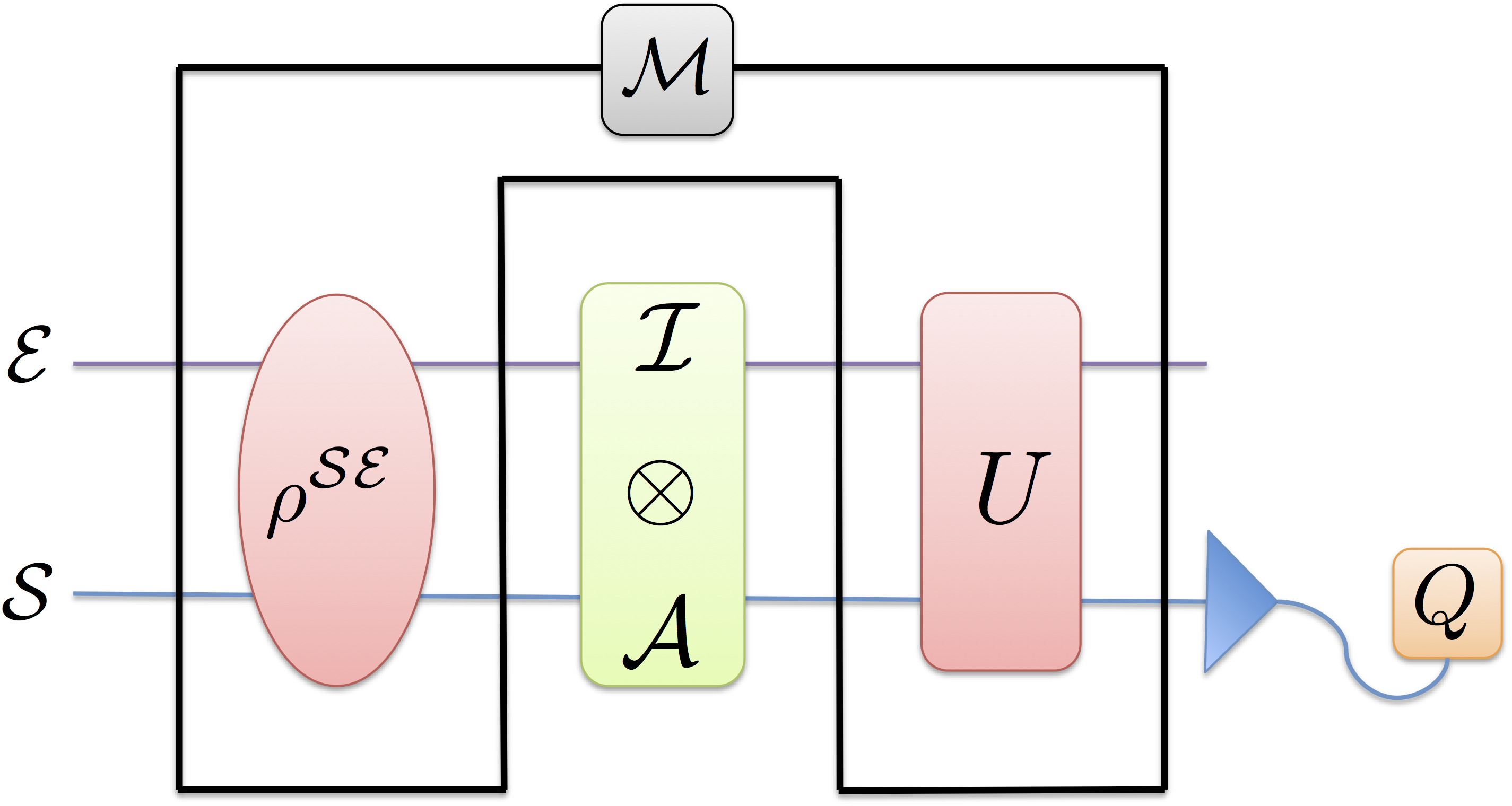}}
Figure 2: \emph{Quantum process tomography with $\M$-map.} At the beginning of the experiment the system-environment state is correlated. A preparation is made on the system and the corresponding output state $Q$ is observed. This process is described by the completely positive map $\M$, which is a function of the initial system-environment state and the unitary dynamics. The $\M$-map takes preparations $\A$ to output states $Q$.
\end{figure}

Note that, in standard quantum process tomography state of $\e$ is a constant of the process, here it is the initial $\se$ state that is the constant of the process, i.e., it is a fixed quantity. Physically, the constancy of $\rse$ means that the experiment should be initialised in the same manner for every run, and then a preparation on $\s$ can be made.

$\M$-map contains both $U$ and $\rse$; however knowing $\M$ is not sufficient to determine $U$ and $\rse$. As expected, it should not be possible to determine $U$ and $\rse$ through measurements and preparations on the system alone without access to the environment. Conversely, $\M$-map contains all information necessary to fully determine the output state for any preparation of $\s$. The advantage of dealing with the $\M$-map is that we have separated the preparation procedure from uncontrollable dynamical elements and the initial conditions. $\M$-map contains all of the dynamical information for the system and in the next section we will extract some of this information from the $\M$-map. First let us mention some properties of $\M$-map derived in Methods: Its action on a mixture of preparations is linear, it preserves trace, it preserves Hermiticity, and it is completely positive.

In Methods we show that $\M-$map can be experimentally determined by making a set of linearly independent preparation of the system. This is similar to what one has to do in standard QPT. In standard QPT a linearly independent set of states are fed into the process and the corresponding outcomes are observed. Knowing the inputs and the outputs the standard process map is determined. The difference here is that a linearly independent set of preparations are fed in to the process. This is of our major result of this paper: We have given a prescription to determine the dynamics of a system  in an operational way, i.e., a mapping from preparations to output states.

\subsection*{Quantifying initial correlations}

The $\M$-map contains the dynamics of the system before any preparation is made on the system. It is a function of the initial state $\rse$ state as well as the $\se$ unitary transformation. $\M$-map is a tensor, taking its trace with respect to the indices that belong to the initial state of $\s$ we can obtain the dynamics of the system as if the initial correlations we absent. Using this with the knowledge of the initial state of $\s$, in Methods we show that from $\M$ we can derive another matrix,
\begin{equation}\label{lmap}
\Lm_{rr'r''; ss's''} = \sum_{\alpha\beta\epsilon} U_{r \epsilon,r' \alpha} \rs_{r''s''} \re_{\alpha\beta} U^*_{s \epsilon,s' \beta}.
\end{equation}
Matrix $\mathcal{L}$ is fully determinable from $\M-$map and the two are the same when there are no initial correlations. We will call the difference between $\M$ and $\Lm$, $\K = \M - \Lm$, the \emph{memory matrix}:
\begin{equation}
\K_{rr'r''; ss's''} = \sum_{\alpha\beta\epsilon} U_{r \epsilon,r' \alpha} \chi^\se_{r''\alpha ; s''\beta} U^*_{s \epsilon,s' \beta} .
\end{equation}
Since $\M$ contains $\rse$ and $\Lm$ contains $\rs \otimes \re$, the difference between the two is a function of only $\chi^\se$. The action of the correlation-memory matrix on a preparation yields
\begin{equation}
\chi^\s_\A(t) =\K(\A) = \tre[U [\A \otimes \I](\chi^\se) U^\dag],
\end{equation}
which is the coherence coming into the system from the initial correlations. For non-Markovian dynamics the future state of $\s$ may depend on the initial $\se$ correlations. This is the non-Markovian `memory' due to the initial  $\se$ correlations and it is a key feature of non-Markovian dynamics~\cite{Rodriguez08}.

The correlation-memory matrix is an important result for studying non-Markovian systems. It is an operational way of measuring the information that flows into $\s$ due to correlations at the time of the preparation. Once $\M-$map is determined, we have the full knowledge of the dynamics of $\s$ that is due to the initial correlations. The correlation-memory matrix provides quantitative information about the initial correlation and it is more than a witness for initial correlations~\cite{Laineetal}.

\subsection*{Operational meaning of not-completely positive maps}

For the special case, when the preparation is chosen to be the identity map, we get pure dynamics of the correlation-memory matrix 
\begin{equation}
\K(\I) = \tre[U \chi^\se U^\dag], 
\end{equation}
which is the reduced dynamics of $\se$ correlations. This is exactly $\B^{\rm aff}$ in Eq.~\ref{CPmap}. From $\M$-map we can determine matrices $\Lm$ and $\K$. In turn, from $\Lm$ we can get $\B^{\rm CP}$ (see Eq.~(\ref{L-cpmap})) and from $\K$ we can get $\B^{\rm aff}$, and together they give us $\B$ of Eq.~(\ref{NCPmap}), which can be a not-completely positive map. This gives not-completely positive maps an operational meaning.

\section*{Discussion}\label{disc}

$\M$-map is the result of a quantum process tomography procedure for initially correlated system-environment states. It is acts on the preparation of the initial state of the system, and only contains dynamical information. We study the properties of $\M$-map, showing it to be linear, preserving of trace and Hermiticity, and completely positive. Dynamical information about the evolution of the initial correlations can be retrieved from $\M-$map, in the form of the correlation-memory matrix $\K$. $\M$-map allows us to determine the output state for any preparation of the system, while the correlation-memory matrix $\mathcal{K}$ provides a quantitative expression for the coherence due to the initial correlations.

An important question is when is $\M$-map relevant? Clearly, when $\s$ and $\e$ are initially uncorrelated then $\mathcal{K}$ will be zero. Alternatively, just the presence of initial $\se$ correlations does not warrant for $\M$-map. Suppose $\chi^\se \ne 0$ but $\tre[U \chi^\se U^\dag] = 0$, then the completely positive map of Eq.~(\ref{CPmap}) would suffice to describe the dynamics correctly for any preparation of $\s$~\cite{modi2012}.

One downside to $\M-$map is that it requires a lot of resources to construct. In standard quantum process tomography $d^2_\s$ input states are fed through the process and the corresponding output states are determined. To determine $\M$ map, $d^4_\s$ preparations are necessary, which is a significant growth over the standard procedure. Therefore an efficient way, such as compressed sensing~\cite{laflamme, arXiv:1205.2300}, to determine this map is desirable. This should be possible, as determining $\M$-map is equivalent to carrying out $d_\s$ standard quantum process tomography procedures.

Another limitation that faces the procedure is the assumption that the preparation acts only on the system and not on the environment. This assumption is crucial, as we are mapping from the set of preparations on the system to the corresponding output states. If this assumption fails, then we would need to make a set of preparations that span the space of operations on the combined system-environment space. However, the environment can be arbitrarily large and we do not have any control over it. Therefore the tools given in this article may not be valid when the preparation affects the environment directly. When the preparation procedure acts on $\e$ as well as $\s$, the positivity of $\M$-map may be affected. Note that, as long the effect of all preparations on $\e$ is a constant for then our prescription remains valid. 

Lastly, since $\M$-map contains all dynamical information, we are able to construct $\B^{\rm CP}$ of Eq.~(\ref{CPmap}) from it. Similarly, from the correlation-memory matrix, we can construct $\B^{\rm aff}$ of Eq.~(\ref{CPmap}). Knowing the two we can determine $\B$ of Eq.~(\ref{NCPmap}), which can be a not-completely positive map. This gives operational meaning to not-completely positive dynamical map as the descriptor for the dynamics of the system when identity preparation is made. On the other hand, the non-completely positive map is not experimentally determinable without determining $\M$-map. Finally, it remains an open question, when $\K > 0$, is $\B$ not completely positive?

\section*{Methods}

The calculations in this sections are done in terms of matrix indices as $\M$-map and the correlation-memory matrix $\K$ are nontrivial tensors. We use the Einstein summation notation, i.e., repeated indices are summed over. Bipartite state of $\se$ is expressed with four indices with the Latin indices belong to $\s$ and greek indices to $\e$. For instance, the state in Eq.~(\ref{corrmatrix}) has the form $\rse_{r\alpha,s\beta} = \rs_{rs} \otimes \re_{\alpha \beta} + \chi^\se_{r\alpha,s\beta}$. A map acting on a density matrix is written as 
$\A_{rr';ss'} \left( \rs_{r's'} \right) = A^k_{rr'} \rs_{r's'} {A^k}^*_{ss'}
= P_{rs}$, where $A^k$ are the Sudarshan-Kraus operators (see~\cite{PhysRev.121.920}). $A^*$ is the complex conjugation of $A$ and $A_{rs} \to A_{sr}$ is the transpose; together they give Hermitian conjugation.

\subsection*{$\M-$map}\label{mmapderi}

Let us rewrite the generalised process equation, Eq.~(\ref{output}), in terms of matrix indices
\begin{equation}
Q_{rs}= U_{r\epsilon;r'\alpha}\A_{r'r'';s's''} \left( \rse_{r''\alpha;s''\beta} \right) U^*_{s\epsilon;s'\beta},
\end{equation}
where the sum over $\epsilon$ is the trace with respect to the environment. We are interested in the reduced dynamics of $\s$ as a function of the preparation procedures. Thus, we can pull the preparation map out of everything else and regard it all as a map acting on the preparation map:
\begin{eqnarray}
Q_{rs}
&=&\nonumber
\left[ U_{r\epsilon;r'\alpha} \rse_{r''\alpha;s''\beta} U^*_{s\epsilon;s'\beta} \right] \left( \A_{r'r'';s's''} \right)\\
&=&\label{RawProcessEquation}
\M_{rr'r'';ss's''} \left( \A_{r'r'';s's''} \right).
\end{eqnarray}
In the last equation, the matrix $\M$ is defined as:
\begin{equation}
\M_{rr'r''; ss's''} =  U_{r \epsilon,r' \alpha} 
\rse_{r''\alpha,s''\beta} U^*_{s \epsilon,s' \beta}.
\end{equation}

\subsection*{Determining $\M-$map}\label{MQPT}

Let $\{ P^\m = \ket{\pi^\m}\bra{\pi^\m} \}$ be a set of pure states that linearly span the space of $\s$. There are $d_\s^2$ such matrices. That is, any state of $\s$ can be written as a linear sum of these pure states: $\rho^\s = \sum_m r_m P^\m$. 

A preparation map acting on $\s$ is a $d^2_\s \times d^2_\s$ Hermitian matrix. Therefore, any matrix in this space can be spanned by a tensor product of the basis matrices $\{ P^\m \otimes P^\n \}$, which is a basis in for $d^2_\s \times d^2_\s$ space of maps. There are $d_\s^4$ elements in the basis $\{ P^\m \otimes P^\n \}$. We can write action of one of these basis element on a density operator on $\s$ as
\begin{equation}\label{linpreps}
\A^\mn(\rse) =\ket{\pi^\n} \braket{ \pi^\m | \rse |\pi^\m } \bra{\pi^\n}.
\end{equation}
It is crucial to note here that $\braket{ \pi^\m |\pi^\n } \ne \delta_{mn}$, as these vectors are eigenvectors of the basis elements $\{ P^\m \}$ that do not commute. 

These preparations are can be thought of as a projection followed by a rotation. Action of any map on space of $\s$ acting on the $\se$ state can be expresses as a linear sum
\begin{eqnarray}
\A ( \rse ) &=& \sum_{mn} \alpha^\mn \A^\mn\\
&=& \sum_{mn} \alpha^\mn \ket{\pi^\n} \braket{ \pi^\m | \rse |\pi^\m } \bra{\pi^\n}\\
&=& \sum_{mn} \alpha^\mn P^\n \otimes \tr_\s \left[ \rse P^\m\right ]\\
&=& \sum_{mn} \alpha^\mn p^\m P^\n \otimes  \rho^{\e|\m},
\end{eqnarray}
where $\alpha^\mn$ are the coefficients that determine $\A$ in terms of $\{\A^\mn\}$. $\rho^{\e|\m}$ is the conditional state of the $\e$ and $p^\m=\tr[ P^\m  \rs ]$ is the probability for the outcome $P^\m$.  

Knowing the output states corresponding to each of these inputs,
\begin{equation}
Q^\mn = \tre[U P^\n \otimes  \rho^{\e|\m} U^\dag],
\end{equation}
along with the success probabilities $p^\m$, for all $m,n$, is enough to predict the output state for any preparation:
\begin{eqnarray}
Q &=& \tre [U\A ( \rse ) U^\dag]\\
&=& \sum_{mn} \alpha^\mn p^\m  \tre [UP^\n \otimes  \rho^{\e|\m} U^\dag]\\
&=& \sum_{mn} \alpha^\mn p^\m  Q^\mn.
\end{eqnarray}
$\M$-map can be determined choosing $\A^\mn$, followed determining the corresponding $Q^\mn$ and $p^\m$, and standard inversion techniques~\cite{Nielsen00a}. Note that any other set of linearly independent preparation can be linearly mapped to the preparations given in Eq.~(\ref{linpreps}), and therefore will suffice.

Determining all $Q^\mn$ is done by quantum state tomography. This is equivalent to carrying out $d_\s$ standard QPT procedures, one each $\rho^{\e|\m}$. Additionally measuring $p^\m$ is equivalent to doing quantum state tomography of $\rs$.

Before moving on a simple example may be useful. For one qubit, we may take the following projectors as a linearly independent basis:
\begin{eqnarray*}
P^{(1)}=\frac{1}{2}(\mathbb{I} + \sigma_1), \;
P^{(2)}=\frac{1}{2}(\mathbb{I} + \sigma_2), \;
P^{(3)}=\frac{1}{2}(\mathbb{I} + \sigma_3), \;
P^{(4)}=\frac{1}{2}(\mathbb{I} - \sigma_1).
\end{eqnarray*}
Note that, this is a linear but not a convex decomposition: $P^{(5)} = \frac{1}{2}(\mathbb{I} + \sigma_2) =P^{(1)} + P^{(4)} - P^{(2)}$. The eigenvectors of $P^{(1)}$, $P^{(2)}$, $P^{(3)}$, and $P^{(4)}$ are $\ket{x+} = \frac{1}{\sqrt{2}} (\ket{0}+\ket{1})$, $\ket{y+} = \frac{1}{\sqrt{2}} (\ket{0}+i\ket{1})$, $\ket{z+} = \ket{0}$, and $\ket{x-} = \frac{1}{\sqrt{2}} (\ket{0}-\ket{1})$ respectively. Using these eigenvectors we can write basis elements for the maps that operate on the space of one qubit. For instance,
\begin{eqnarray*}
&& \A^{(1,1)}(\rse) = \ket{x+} \bra{x+} \rse \ket{x+} \bra{x+} = p^{(x+)} \ket{x+} \bra{x+} \otimes \rho^{\e|(x+)},\\
&& \A^{(3,4)}(\rse) = \ket{x-} \bra{z+} \rse \ket{z+} \bra{x-} = p^{(z+)} \ket{x-} \bra{x-} \otimes \rho^{\e|(z+)},\\
&& \A^{(4,2)}(\rse) = \ket{y+} \bra{x-} \rse \ket{x-} \bra{y+}= p^{(x-)} \ket{y+} \bra{y+} \otimes \rho^{\e|(x-)},
\end{eqnarray*}
and so on.

\subsection*{Detecting initial correlations}\label{inicorr}

The initial state of the system is labeled by indices $r''$ and $s''$. Tracing over everything else we can find the initial state of $\s$ (before preparation) from $\M$-map:
\begin{equation}
\frac{1}{d_\s} \delta_{rs}\delta_{r's'} \M_{rr'r'';ss's''}
=\rho^\mathcal{S}_{r''s''}.\label{initialstates}
\end{equation}
This is, of course, attainable by doing state tomography at the beginning of the experiment, by measuring the values of $p^\m$ from last section.

Next, let us the trace over the system indices $r''$ and $s''$
\begin{equation}\label{processfromM}
\delta_{r''s''} \M_{rr'r''; ss's''} =  U_{r \epsilon,r' \alpha} \re_{\alpha\beta} U^*_{s \epsilon,s' \beta}
= \B^{\rm CP}_{rr';ss'}.
\end{equation}
The last equation is exactly the dynamical map in the absence of initial correlations, given in Eq.~(\ref{CPmap}). In other words, in the absence of initial correlations, QPT would yield this map.

This means, even though the $\M$-map contains the information about uncorrelated $\se$ state and the correlations separately. Consider the following matrix composed of the matrices in Eqs.~(\ref{initialstates}) and (\ref{processfromM})
\begin{eqnarray}
\mathcal{L}_{rr'r'';ss's''}
&=&\B^{CP}_{rr';ss'}\rs_{r''s''}\nonumber\\
&=& U_{r \epsilon,r' \alpha} \rs_{r''s''}\re_{\alpha\beta} U^*_{s \epsilon,s' \beta}.\label{L-cpmap}
\end{eqnarray}
The last equation is similar to the expression for the $\M$-map, except the state of the system and the state of the environment are uncorrelated.

Writing the state of $\se$ in $\M$-map in terms of Eq.~(\ref{corrmatrix}), we get
\begin{equation}
\M^{(rs)}_{r'r'';s's''}
= U_{r \epsilon,r' \alpha} 
(\rs_{r''s''}\re_{\alpha\beta} +\chi^{\se}_{r''\alpha;s''\beta})
U^*_{s \epsilon,s' \beta}.
\end{equation}

Now we can define the correlation-memory matrix as
\begin{eqnarray}
\K_{rr'r'';ss's''} &=&\M_{rr'r'';ss's''}- \Lm_{rr'r'';ss's''}\\
&=& U_{r \epsilon,r' \alpha} \chi^{\se}_{r''\alpha;s''\beta} U^*_{s \epsilon,s' \beta}.\label{memoryeq}
\end{eqnarray}

\subsection*{Properties of $\M$}\label{app1}

\subsubsection*{Linearity}

Mathematically, $\M-$map acts on the preparation map just as the dynamical map acts on a density operator. In fact, we are not varying the initial state of the system, rather the preparation procedure on that state. Therefore the linearity of quantum mechanics is preserved for the $\M$-map acting on different preparation procedures, i.e.
\begin{equation}
\M\left[\alpha_1 \A^{(1)} + \alpha_2 \A^{(2)}\right] = \alpha_1 \M\A^{(1)} + \alpha_2 \M\A^{(2)}.
\end{equation}
This is very much like the dynamical maps action on mixtures of states. Furthermore, if we show that the $\M-$map preserves trace, Hermiticity, and positivity on its domain then all of these properties will be preserved on the state space. In other words for any preparation, $\A^\m$ that preserves trace, Hermiticity, and positivity, the action of the $\M-$map on it will yield an output state, $Q^\m$, that is unit-trace, Hermitian and positive.

\subsubsection*{Trace preservation}

Let us start with the trace of $\M$ with respect to the final indices $r$ with $s$:
\begin{eqnarray}
\tr_{(rs)}[\M] &=&\delta_{rs} \M_{rr'r'';ss's''}\nonumber\\
&=& U_{r \epsilon ,r' \alpha} \rse_{r''\alpha ,s''\beta}U^*_{r \epsilon ,s' \beta}.
\end{eqnarray}
Since $U^\dag U=\sum_{r\epsilon} U^*_{r\epsilon,s'\beta} U_{r\epsilon,r'\alpha} =\mathbb{I}$, then
\begin{equation}
\tr_{(rs)} [\M] = \delta_{r's'} \delta_{\alpha \beta} \rse_{r''\alpha,s''\beta} = \mathbb{I} \otimes \rs.
\end{equation}
A preparation acting on the above matrix will yield  
\begin{equation}
\tr_{(rs)} [\M](\A)=\tr[\A(\rs)]=1.
\end{equation}
The implication being $\M$ preserves the trace of $\A(\rs)$. As long as the preparation is trace a preserving operation we get a unit-trace matrix for the output state.

\subsubsection*{Hermiticity preservation}

As with the case of general quantum operations, matrix $\M$ is Hermitian. This is easy to see by taking the complex conjugate of matrix $\M$,
\begin{eqnarray}
\left(\M_{rr'r'';ss's''}\right)^* &=&\sum_{\alpha\beta\epsilon}  U_{s \epsilon,s' \beta} \rse_{s''\beta ,r''\alpha} U^*_{r \epsilon,r' \alpha,} \nonumber\\
&=&\M_{ss's'';rr''r'}.
\end{eqnarray}
The complex conjugate of $\M$ is not only the transpose of $\M$, but each element of $\M$ is also transposed.  Hence $\M$ is a Hermitian matrix.

\subsubsection*{Positivity of $\M$-map}

The $\M$-map is composed of a unitary matrix operating on a density matrix.  Then we can take the square root of the density matrix to get
\begin{eqnarray}
\M_{rr'r'';ss's''} &=& U_{r \epsilon,r' \alpha} \sqrt{\rse}_{r''\alpha,\sigma \gamma} \sqrt{\rse}_{\sigma \gamma,s''\beta} U^*_{s \epsilon,s' \beta}\nonumber\\
&=& M^{\mu}_{r;r'r''} {M^{{\mu}^*}_{s's'';s}},
\end{eqnarray}
where $M=U \sqrt{\rse}$ and $\mu=\sigma\gamma\epsilon$.  We have written the $\M$-map in operator sum representation, hence it is completely positive.  Where $M^{\mu}$ are the Sudarshan-Kraus operators~\cite{PhysRev.121.920, kraus}. This means, the $\M$-map acting on any preparation procedure will lead to a physical state. This was not the case when a standard QPT procedure is carried out on initially correlated $\se$ states. The action of $\M$-map can now be written as 
\begin{equation}
\M(\A)=	\sum_{\mu}	M^{\mu}	\A	\; M^{\mu^\dag}.
\end{equation}

The properties shown above are precisely the conditions for a generic quantum operation to preserve trace, Hermiticity and positivity. Therefore $\M-$map preserves the attributes on the preparations, which in return will preserve these attributes on the states.

\bibliographystyle{unsrt}

\section*{Acknowledgments} 

We are grateful to A. Brodutch, A. Rezakhani, C. A. Rodr\'iguez-Rosario, and Keith Burnett for valuable conversations. We acknowledge the financial support of John Templeton Foundation, National Research Foundation, and Ministry of Education in Singapore. Part of the work presented here was done while the author was at the University of Texas at Austin.





\end{document}